\documentclass[a4paper,twocolumn,prl,showpacs,amsmath,amssymb,superscriptaddress]{revtex4}

\usepackage[T1]{fontenc}
\usepackage[latin1]{inputenc}
\usepackage{color}
\usepackage{graphicx}
\usepackage{txfonts}

\DeclareMathAlphabet{\mathbfsf}{\encodingdefault}{\sfdefault}{bx}{n}

\renewcommand{\vec}[1]{{\ensuremath\mathbf{#1}}}
\newcommand{\tens}[1]{\ensuremath\mathbfsf{#1}}
\newcommand{\grad}[1]{\vec{\nabla}{#1}}
\newcommand{\curl}[1]{\vec{\nabla}\times{#1}}
\renewcommand{\div}[1]{\vec{\nabla}\cdot{#1}}
\newcommand{\f}[2]{\frac{#1}{#2}}

\newcommand{\diff}{d}
\newcommand{\dpart}[2]{\f{\partial {#1}}{\partial {#2}}}
\newcommand{\dpartshort}[2]{\partial {#1}/\partial {#2}}
\newcommand{\dpartsupershort}[2]{\partial_{#2} {#1}}
\newcommand{\dlag}[2]{\f{d #1}{d #2}}
\newcommand{\dlagshort}[2]{d {#1}/d {#2}}

\newcommand{\Equ}[1]{Equation~({\ref{#1}})}

\newcommand{\equ}[1]{Eq.~({\ref{#1}})}
\newcommand{\equs}[2]{Eqs.~({\ref{#1}})-(\ref{#2})}
\newcommand{\equsthree}[3]{Eqs.~({\ref{#1}})-(\ref{#2})-(\ref{#3})}
\newcommand{\eps}{\ensuremath\varepsilon}

\newcommand{\rhoi}{\rho_i}

\newcommand{\vthi}[0]{v_{\mathrm{th}i}}
\newcommand{\vthiperp}[0]{v_{\mathrm{th}i\perp}}

\newcommand{\kpar}[0]{k_{\parallel}}
\newcommand{\kperp}[0]{k_{\perp}}
\newcommand{\vkperp}[0]{\vec{k}_{\perp}}

\newcommand{\vvperp}[0]{\vec{v}_{\perp}}

\newcommand{\vperp}[0]{v_{\perp}}
\newcommand{\vpar}[0]{v_{\parallel}}
\newcommand{\vparEmu}[0]{\sqrt{2(E\!-\!\mu B)}}

\newcommand{\gradpar}[1]{\grad{}_\parallel{\,#1}}
\newcommand{\nablapar}[1]{\nabla_\parallel{\,#1}}

\newcommand{\nablaparsquare}[1]{\nabla_\parallel^2{#1}}
\newcommand{\nablaperpsquare}[1]{\nabla_\perp^2{#1}}
\newcommand{\gradperp}[1]{\grad{}_\perp{#1}}

\newcommand{\curlperp}[1]{\gradperp{}\times{\,#1}}

\newcommand{\nuii}{\nu_{ii}}
\newcommand{\nuiieff}{\nu_{ii,\mathrm{eff}}}
\newcommand{\omegab}{\omega_B}
\newcommand{\bo}{{B_0}}

\newcommand{\vbo}{{\vec{B}_0}}
\newcommand{\dB}{{\delta B}}
\newcommand{\dvB}{{\delta \vec{B}}}
\newcommand{\bhat}{{\hat{\vec{b}}}}
\newcommand{\bhato}{{\hat{\vec{b}}_0}}

\newcommand{\bfourpar}{{\delta B_4^\parallel}}

\newcommand{\bfourparkhat}{{\delta\hat{B}_{4\vec{k}}^\parallel}}
\newcommand{\bfourpartilde}{{\delta\tilde{B}_4^\parallel}}

\newcommand{\bfiveperp}{{\delta \vec{B}_5^{\perp}}}
\newcommand{\bsixpar}{{\delta B_6^{\parallel}}}

\newcommand{\uzeroi}{{\vec{u}_{0i}}}
\newcommand{\dvui}{\delta \vec{u}_i}

\newcommand{\ufiveperpi}{{\delta \vec{u}_{5i}^{\perp}}}

\newcommand{\ppari}{{p_{i}^{\parallel}}}
\newcommand{\pperpi}{{p_{i}^{\perp}}}
\newcommand{\pzeropari}{{p_{0i}^{\parallel}}}
\newcommand{\pzeroperpi}{{p_{0i}^{\perp}}}
\newcommand{\ptwoperpi}{{p_{2i}^{\perp}}}

\newcommand{\pfourperpi}{{\delta p_{4i}^{\perp}}}
\newcommand{\pfourpari}{{\delta p_{4i}^{\parallel}}}

\newcommand{\pfiveperppari}{{\delta \vec{p}_{5i}^{\perp\parallel}}}

\newcommand{\psixperpperpi}{{\delta \tens{P}_{6i}^{\perp\perp}}}

\newcommand{\psixperpir}{{\delta p_{6i}^{\perp}\phantom{}^{\mathrm{(res)}}}}
\newcommand{\psixperpirtilde}{{\delta \tilde{p}_{6i}^{\perp}\phantom{}^{\mathrm{(res)}}}}

\newcommand{\fzeroi}{f_{0i}}
\newcommand{\ftwoi}{f_{2i}}
\newcommand{\df}{\delta f}
\newcommand{\ffouri}{\delta f_{4i}}
\newcommand{\ffivei}{\delta f_{5i}}
\newcommand{\fsixi}{\delta f_{6i}}

\newcommand{\ffourir}{\delta f_{4i}^{\mathrm{(res)}}}
\newcommand{\ffourirtilde}{\delta \tilde{f}_{4i}^{\mathrm{(res)}}}
\newcommand{\ffourikhatr}{\delta {\hat{f}_{4i\vec{k}}^{\mathrm{(res)}}}}

\newcommand{\betazeroperpi}{{\beta_{0i}^{\perp}}}

\newcommand{\vtilde}{v_*}
\newcommand{\rhotilde}{\rho_*}

\newcommand{\xitr}{{\xi_{\mathrm{tr}}}}

\newcommand{\lin}{{\mbox{\tiny $L$}}}
\newcommand{\nonlin}{{\mbox{\tiny $NL$}}}
\newcommand{\gammalin}{\gamma_{\lin}}
\newcommand{\gammanl}{\gamma_{\nonlin}}

\newcommand{\lineav}[1]{\overline{#1}}
\newcommand{\bounceav}[1]{\left<{#1}\right>}
\newcommand{\bigO}[1]{O(#1)}

\begin{document}

\title{Nonlinear mirror instability}

\author{F. Rincon}
\email{francois.rincon@irap.omp.eu}

\affiliation{Universit\'e de Toulouse; UPS-OMP; IRAP; Toulouse, France}
\affiliation{CNRS; IRAP; 14, avenue Edouard Belin, F-31400 Toulouse, France}
\author{A. A. Schekochihin}
\affiliation{Rudolf Peierls Centre for Theoretical Physics, University of
Oxford, 1 Keble Road, Oxford, OX1 3HQ, United Kingdom}
\affiliation{Merton College, Oxford OX1 4JD, United Kingdom}
\author{S. C. Cowley}
\affiliation{{CCFE}, Culham Science Centre, Abingdon, OX14 3DB, United Kingdom}
\affiliation{Blackett Laboratory, Imperial College, Prince Consort
  Road, London, SW7 2BZ, United Kingdom}

\date{\today}

\begin{abstract}
Slow dynamical changes in magnetic-field strength and invariance of
the particles' magnetic moments generate ubiquitous pressure
anisotropies in weakly collisional, magnetized astrophysical plasmas.
This renders them unstable to fast, small-scale mirror and firehose
instabilities, which are capable of exerting feedback on
the macroscale dynamics of the system. By way of a new asymptotic
theory of the early nonlinear evolution of the mirror instability in a
plasma subject to slow shearing or compression, we show that the instability does not
saturate quasilinearly at a steady, low-amplitude level. Instead, the
trapping of particles in small-scale mirrors leads to nonlinear secular
growth of magnetic perturbations, $\dB/B \propto t^{2/3}$. Our
theory explains recent collisionless simulation results, provides a
prediction of the mirror evolution in  weakly collisional plasmas and
establishes a foundation for a theory of nonlinear mirror
dynamics with trapping, valid up to $\dB/B =\bigO{1}$.
\end{abstract}


\maketitle

\paragraph{Introduction.}
Dynamical, weakly collisional {high-$\beta$} plasmas develop
pressure anisotropies with respect to the magnetic field as a result
of the combination of slow changes in magnetic-field strength $B$
and conservation of the first adiabatic invariant of particles
$\mu=\vperp^2/2B$. This renders them unstable to fast (ion cyclotron
timescale  $\Omega_i^{-1}$), small-scale (ion gyroscale $\rhoi$)
firehose, mirror and ion cyclotron instabilities
\citep{rosenbluth56,chandra58,parker58,vedenov58,rudakov61,gary92},
whose observational signatures have been
reported in the solar wind \citep{hellinger06,bale09} and
planetary magnetosheaths
\citep{kaufmann70,erdos96,andre02,joy06,genot09,horbury09,soucek11}. 
These instabilities are also thought to be excited in energetic
astrophysical environments such as the intracluster medium (ICM)
\citep{fabian94,carilli02,govoni04,schekochihin05,peterson06}, the
vicinity of accreting black holes
\citep{quataert01,narayan05,blaes13}, or the warm ionized interstellar
medium \citep{hall80, ferriere01}, producing strong dynamical
feedback at macroscales, with critical astrophysical implications
\citep{chandran98,sharma06,sharma07,kunz11,schekochihin06,mogavero14}.
A self-consistent description of the multiscale physics of such plasmas
requires understanding how these instabilities saturate nonlinearly.

Let us consider a typical situation in which slow changes in $B$ due
to shearing, compression or expansion of the plasma at large
(``fluid'') scales build up ion pressure anisotropy $\Delta_i\equiv
(\pperpi-\ppari)/\pperpi$, driving the plasma through
either the ion firehose instability
boundary ($\Delta_i < -2/\beta_i$, with $\beta_i=8\pi p_i/B^2$)  in
regions of decreasing field, or the mirror instability boundary
 ($\Delta_i \gtrsim 1/\beta_i$) in regions of increasing field. This triggers
exponential growth on timescales up to $\Omega_i^{-1}$, much faster
than the shearing timescale $S^{-1}$. The separation between these
timescales implies that the instabilities always operate close to
threshold and regulate the levels of pressure anisotropy in the plasma
nonlinearly. However, how they achieve that in the face of the
slowly changing $B$ constantly generating more pressure anisotropy,
remains an open question.

In the simplest case of the parallel firehose instability, the growth
of magnetic perturbations $\dvB$ leads to an increase of the average
(r.m.s.)~field strength and perpendicular pressure, which drives the
anisotropy back to marginality,
$\Delta_i(t)\rightarrow-2/\beta_i$. If a weakly unstable initial
state $\Delta_{io}-2/\beta_i<0$ is postulated with no further driving
of $\Delta_i$, quasilinear theory \citep{shapiro64} predicts
saturation at a steady, low amplitude $\dB/B \sim
\vert\Delta_{io}+2/\beta_i\vert^{1/2}\ll 1$. However, the shearing or
expansion process that drove the plasma through the instability
boundary in the first place, must ultimately become important again
once quasilinear relaxation has pushed the system sufficiently close
back to marginality. When such continued driving is accounted for,
asymptotic theory \citep{schekochihin08,rosin11} predicts secular
growth of perturbations as $\dB/B\propto t^{1/2}$ up to
$\dB/B=\bigO{1}$ (cf.~\cite{matteini06}), a very different
outcome from steady-state, low-amplitude saturation.

The nonlinear dynamics of the mirror instability
\citep{tajiri67,hasegawa69,southwood93,hellinger07} in a weakly collisional
shearing (or compressing) plasma driven through its instability
boundary is much more involved and has only recently been explored
numerically \citep{kunz14a,riquelme14}. In this Letter, we show that weakly
nonlinear mirror modes in such conditions do not saturate
quasilinearly at a steady, low amplitude either, but continue to grow
secularly as $\dB/B \propto t^{2/3}$. To do this, we introduce
a new asymptotic theory in the spirit of earlier theoretical work
\citep{califano08,schekochihin08,rosin11}, in which the combined
effects of weak collisionality, large-scale shearing, quasilinear relaxation
\citep{shapiro64,pokhotelov08,hellinger09}, particle trapping
\citep{kivelson96,pantellini98,istomin09,pokhotelov10} and
 finite ion Larmor radius (FLR) are all self-consistently retained.

\paragraph{Asymptotic theory.} 
We consider the simplest case of a plasma consisting of cold electrons
\footnote{Formally, we first expand the electron Vlasov equation in
  $\sqrt{m_e/m_i}$ and  then take the limit $T_e\ll T_i$, so electrons
  still stream along the field very fast. This is purely  a
  matter of analytical convenience: our results also hold for
  hot, isothermal electrons.}
and hot ions of mass $m_i$, charge $q_i=Ze$, and 
thermal velocity $\vthi=\sqrt{2\,T_i/m_i}$, coupled 
to the electromagnetic fields $\vec{E}$ and $\vec{B}$. The dynamics 
is governed by the non-relativistic Vlasov-Maxwell system, 
\begin{equation}
  \label{eq:vlasov}
  \dpart{f_i}{t}+\vec{v}\cdot\grad{f_i}+\f{q_i}{m_i}\left(\vec{E}+\f{\vec{v}\times\vec{B}}{c}\right)\cdot\dpart{f_i}{\vec{v}}=C\left[f_i\right]\,,
\end{equation}
\begin{equation}
  \label{eq:maxwell}
  \div{\vec{B}}=0~,\quad \dpart{\vec{B}}{t}=-c\,\curl{\vec{E}}~,\quad
  \vec{j}=\f{c}{4\pi}\curl{\vec{B}}~,
\end{equation}
and Ohm's law describing the force balance for electrons,
\begin{equation}
\label{eq:ohm}
\vec{E}+\f{\vec{u}_i\times\vec{B}}{c}
=\f{\left(\curl{\vec{B}}\right)\times\vec{B}}{4\pi
      Zen_i}~.
\end{equation}
Here, $f_s(t,\vec{r},\vec{v})$, $n_s(t,\vec{r})=\int f_s\,\diff^3\vec{v}$ and
$\vec{u}_s(t,\vec{r})=\int\vec{v}f_s\,\diff^3\vec{v}$ are, respectively,
the distribution function, number density and mean velocity of species
$s=(i,e)$, $\vec{j}=e \,n_e\left(\vec{u}_i-\vec{u}_e\right)$
is the total current density given the quasineutrality condition
$n_e=Z n_i$. In the following, we use the ion peculiar velocity
$\vec{v}'=\vec{v}-\vec{u}_i$ as the velocity-space variable and will
henceforth drop the primes. Taking the first moment of
\equ{eq:vlasov} and using \equs{eq:maxwell}{eq:ohm}, we obtain the ion
momentum equation 
\begin{equation}
\label{eq:momentumi}
\dlag{\,\vec{u}_i}{t} =
-\f{\div{\tens{P}_i}}{m_in_i}+\f{\left(\curl{\vec{B}}\right)\times\vec{B}}{4\pi
  m_in_i}~,
\end{equation}
where $\dlagshort{}{t}=\dpartshort{}{t}+\vec{u}_i\cdot\grad{}$ and
$\tens{P}_i=m_i\int \vec{v}\vec{v}f_i\,\diff^3\vec{v}$ is
the ion pressure tensor. Introducing $\bhat\equiv\vec{B}/B$, using
\equs{eq:maxwell}{eq:ohm}, we obtain the evolution equation
for the field strength:
\begin{equation}
  \label{eq:induction}
\dlag{\ln
  B}{t}=\bhat\bhat:\grad{\vec{u}_i}-\div{\vec{u}_i}-\f{\bhat}{B}\cdot\curl{\left(\f{\vec{j}\times\vec{B}}{Zen_i}\right)}~.
\end{equation}

Our derivation is based on an asymptotic expansion of these
equations. The separation between the
slow magnetic-field-stretching timescale and the
fast instability timescale implies that the distance to instability
threshold $\Gamma \sim \Delta_i - 1/\beta_i$ must remain small, which
provides us with a natural expansion parameter. In order to
study the dynamics in this regime, we start from an already weakly
unstable situation and order $\Gamma=\bigO{\eps^2}$, with $\eps \ll
1$. We then construct a ``maximal'' ordering (summarized in
\equs{eq:expansionsummaryfirst}{eq:expansionsummarylast})
retaining ion FLR, collisional, quasilinear and trapping effects, as
well as the effect of continued slow shearing. Following
\cite{hellinger07}, we order the time and spatial
scales of mirror modes as $\gamma\sim\eps^2\kpar\vthi$, $\kperp \sim
\eps^{-1}\kpar$, $\rho_i^{-1}\sim\eps^{-2}\kpar$,
where $\gamma$ is the instability growth rate, $(\vkperp,\kpar)$
the typical perturbation wavenumbers (defined with respect
to the unperturbed field), $\rhoi=\vthiperp/\Omega_i$,
and $\Omega_i^{-1}=(m_ic)/q_i B\sim\eps^{-2}\kpar\vthi$. The ion
distribution function is expanded as $f_i=\fzeroi+\ftwoi+\df$, 
where $\fzeroi$ provides the required pressure anisotropy to pin
the system at the threshold, $\ftwoi$ provides an
extra $O(\eps^2)$ anisotropy to drive the system away from it, 
and $\df$ contains mirror perturbations. We also expand
$\vec{B}=\vbo+\dvB$ and $\vec{u}_i=\uzeroi+ \dvui$, where
$\vbo$ and $\uzeroi$ have no instability-scale variations, $\uzeroi$
is the slow, large-scale shearing/compressive motion, and
$\dvB$ and $\dvui$ are the mirror perturbations. 

The ordering of $\df$, $\dvB$, $\dvui$ and of the remaining timescales is
guided by physical considerations. The critical pitch-angle
$\xi=\vpar/v$ below which particles get trapped by magnetic
fluctuations is $\xitr=\left(\dB/\bo\right)^{1/2}$ and the
corresponding bounce frequency is $\omegab\sim\kpar\vthi\,\xitr$.
To retain their contribution in our calculation, we order
$\omegab\sim\gamma\sim\eps^2\kpar\vthi$, which provides us
with the ordering $\dB/\bo=\bigO{\eps^4}$  (plus higher-order
terms). For consistency of the $\eps$ expansion, we must order
$\df=\bigO{\eps^4}$, $\dvui=\bigO{\eps^5}$ and higher. Averaging
\equ{eq:induction} over instability scales and ignoring quadratic
nonlinearities,  we obtain 
\begin{equation}
\label{eq:shear}
\dlag{\ln\bo}{t} = \bhato\bhato:\grad{\uzeroi} -\div{\uzeroi}\,\,
\equiv\,\, S~.
\end{equation}
Subtracting \equ{eq:shear} from \equ{eq:induction}, we find that
\begin{equation}
\label{eq:inductionpert}
\dlag{}{t}\f{\dB}{\bo}= \bhato\bhato:\grad{\dvui}
\end{equation}
to all relevant orders \footnote{A Hall term
 $-\left(\rhoi\vthiperp/\betazeroperpi\right)\bhato\cdot\left[\curlperp{\left(\nablapar{\bfiveperp/\bo}\right)}\right]$
 is formally present in the induction equation at the maximum order
 $\bigO{\eps^6}$ needed here. However, it can be proven to be zero
 because of the particular polarization of mirror modes
 \citep{califano08} and has therefore been summarily
 discarded in \equ{eq:inductionpert} to simplify the algebra.}.
We order $S\equiv\dlagshort{}{t}\,(\ln \bo)$ the same size as
$\dlagshort{}{t}\,(\delta
B/\bo)\sim \eps^6\kpar\vthi$ so as to be able to investigate how a
slow change in field strength affects the dynamics.
The aforementioned timescale separation $S/\Omega_i$
is now related to $\eps$ through $\eps \sim (S/\Omega_i)^{1/8}$ 
($\eps\sim 0.01$ for the ICM \citep{rosin11}).

A separate asymptotic treatment is required for low-pitch-angle
resonant particles, which develop a velocity-space
boundary layer and evolve into a separate population of trapped
particles on the instability timescale (the process is reminiscent of
``nonlinear Landau damping'' \citep{istomin09,dawson61,oneil65}).
As can be seen by considering a simple Lorentz pitch-angle scattering
operator \citep{helander02},
$C\left[f_i\right]=(\nuii/2)\,\dpartsupershort{}{\xi}\left[\left(1-\xi^2\right)\dpartsupershort{f_i}{\xi}\right]$,
this results in a boost of their effective collisionality, 
$\nuiieff\sim\nuii/\xitr^2\gg \nuii$ for $\xi<\xitr\ll 1$.
To retain this effect, we order
$\nuiieff\sim\gamma\sim\omegab$, or $\nuii\sim
\kpar\vthi\left(\dB/\bo\right)^{3/2}\sim\eps^6
\kpar\vthi$ (this preserves the low-collisionality
assumption in the sense that the rest of the distribution
relaxes on a timescale $1/\nuii \gg 1/\gamma$). 
The maximal mirror ordering is summarized as follows:
\begin{eqnarray}
& f_i= \fzeroi+ \ftwoi+ \ffouri+\cdots,\label{eq:expansionsummaryfirst}\\
& \gamma\sim\eps^2\kpar\vthi~,\ \Omega_i\sim\eps^{-2}\kpar\vthi~,\
 S\sim\nuii\sim\eps^6\kpar\vthi~,\\
& \rho_i^{-1}\sim\eps^{-2}\kpar~,\ \kperp \sim  \eps^{-1}\kpar~, \\
& \vec{B}= \vbo + \bfourpar\bhato + \bfiveperp +\cdots,\,\ \vec{u}_i = \uzeroi + \ufiveperpi +\cdots, \\
& \xitr \sim \left(\bfourpar/\bo\right)^{1/2}\sim \eps^2,\,\ \omegab
\sim \nuiieff\sim \gamma \sim \eps^2\kpar\vthi~.\label{eq:expansionsummarylast}
\end{eqnarray}

Taking the three lowest non-trivial orders of \equ{eq:vlasov}, we
first find that $\fzeroi$, $\ftwoi$, $\ffouri$ are gyrotropic. 
Expanding and gyroaveraging \equ{eq:vlasov} up to
$\bigO{\eps^4}$ then gives $\ffouri$ in terms of the mirror
perturbation $\bfourpar/\bo$, from which the perturbed scalar
pressures  $\pfourperpi$ and $\pfourpari$
are derived (note that resonant/trapped  particles are not involved
at this stage). Taking the perpendicular projection of \equ{eq:momentumi} at
the lowest order $\bigO{\eps^3}$, we obtain the threshold condition
for the mirror instability \citep{hellinger07}:
\begin{equation}
\label{eq:Gamma0}
\Gamma_0=
-\f{2\,m_i}{\pzeroperpi}\int\f{\vperp^4}{4}\left.\dpart{\fzeroi}{\vpar^2}\right|_{\vperp}\!\!\!\!\diff^3\vec{v}-\f{2}{\betazeroperpi}-2=0~.
\end{equation}
Next, we expand and gyroaverage \equ{eq:vlasov} to three further
orders, up  to $\bigO{\eps^7}$. This tedious calculation yields FLR 
corrections and resonant effects (not shown, see
\cite{califano08} for an almost identical procedure) and provides
us with explicit expressions for $\ffivei$ and $\fsixi$, from which
we obtain higher-order elements of $\tens{P}_i$,  $\pfiveperppari
\equiv m_i\int \vvperp\vpar\, \ffivei\,\diff^3\vec{v}$ and
$\psixperpperpi$, in terms of $\bfourpar$ and $\bsixpar$. No new
information arises from \equ{eq:momentumi} at $\bigO{\eps^4}$.
Using these results and \equ{eq:Gamma0} in the perpendicular
projection of \equ{eq:momentumi} at $\bigO{\eps^5}$, 
we derive the pressure balance condition:
\begin{equation}
  \label{eq:momentumiorder5perp}
\left[\Gamma_2+\f{3}{2}\,\rhotilde^2\nablaperpsquare{}
-\left(\f{\pzeroperpi-\pzeropari}{\pzeroperpi}+\f{2}{\betazeroperpi}\right)
\f{\nablaparsquare{}}{\nablaperpsquare{}}\right]\gradperp{\f{\bfourpartilde}{\bo}}=
\displaystyle{\gradperp{\f{\psixperpirtilde}{\pzeroperpi}}}~,
\end{equation}
where the resonant/trapped particle pressure is
\begin{equation}
\label{eq:p6res}
\psixperpirtilde = m_i\int_{|\xi|<\xitr}\!\!\!
\f{\vperp^2}{2}\, \ffourirtilde\,\diff^3\vec{v}~,
\end{equation}
$\ffourir\!$ is the resonant part of the perturbed distribution
function, the second-order distance to instability threshold is
\begin{equation}
\label{eq:Gamma2}
\Gamma_2 = -\,\f{2\,m_i}{\pzeroperpi}\left(\int\f{\vperp^4}{4}\left.\dpart{\ftwoi}{\vpar^2}\right|_{\vperp}\!\!\!\!\diff^3\vec{v}
+\!\!\int_{|\xi|<\xitr}\!\!\!\!\f{\vperp^4}{4}\left.\dpart{\overline{\ffourir}}{\vpar^2}\right|_{\vperp}\!\!\!\!\diff^3\vec{v}\right)-\f{2\,\ptwoperpi}{\pzeroperpi}~,
\end{equation}
and the effective Larmor radius is
\begin{equation}
\rhotilde^2 = \f{\rho_i^2}{12}\,\f{m_i}{\pzeroperpi\,\vthiperp^2}\int\left(-\vperp^6\left.\dpart{\fzeroi}{\vpar^2}\right|_{\vperp}-3\,\vperp^4\,\fzeroi\right)\,\diff^3\vec{v}~.
\end{equation}
Tildes denote fluctuating (zero field-line average)
parts of the perturbed fields and overlines denote line averages. 
The l.h.s.~of \equ{eq:momentumiorder5perp} describes the non-resonant 
response \footnote{Nonlinearities quadratic in $\dB/B$ are
  negligible here because of the weakly nonlinear ordering
  $\dB/B\sim\Gamma^2$, which differs from $\dB/B\sim\Gamma$  used in
  \cite{califano08}.}. $\Gamma_2$ and $\psixperpirtilde$ depend on
the regime considered. However, both only involve $\ffourir$
because restricting the integration to $\xitr$ brings in an extra
$\bigO{\eps^2}$ smallness, whereas FLR corrections only start to
affect the distribution function at $\bigO{\eps^6}$ within our expansion. 
Thus, $\ffourirtilde$ and $\psixperpirtilde$ can
be directly calculated from the much simpler drift-kinetic equation
which, in $(\mu,\vpar)$ variables, reads \citep{kulsrud83}:
\begin{equation}
  \label{eq:driftkin}
  \dlag{f_i}{t}+\vpar\,\gradpar f_i= - \mu B\, \left(\div{\bhat}\right)\,\dpart{f_i}{\vpar}+C[f_i]
\end{equation}
to all orders relevant to our calculation (here $E_\parallel=0$
because the electrons are cold). For resonant
particles, $\vpar\sim\eps^2\vthi$ and $\dpartshort{\ffourir}{\vpar}
\sim (\eps^{-2}/\vthi)\,\ffourir$, so the expansion of
\equ{eq:driftkin} at the first non-trivial order $\bigO{\eps^6}$ is
\begin{eqnarray}
  \dlag{\ffourir}{t}+\vpar\,\gradpar \ffourir  & = &  \mu
  \bo\,\left(\f{\gradpar{\bfourpar}}{\bo}\right)\left(\dpart{\fzeroi}{\vpar}+\dpart{\ffourir}{\vpar}\right)
\nonumber \\
&&
+C[\ffourir] \label{eq:driftkinexpand}~.
\end{eqnarray}
We have omitted the collision term $C[\fzeroi]$: it gives a 
$\bigO{\eps^4}$ correction to the line-averaged pressure on
the instability timescale that can be absorbed into $\Gamma_0$.

\paragraph{Linear and quasilinear regimes.} Neglecting the nonlinear
and collision terms in \equ{eq:driftkinexpand}  and taking its
space-time Fourier transform, we obtain the linear solution:
\begin{equation}
  \label{eq:f4itilde}
  \ffourikhatr = \f{\mu\,i\kpar\,\bfourparkhat}{\gammalin+ i\,\kpar\vpar}\dpart{\fzeroi}{\vpar}~,
\end{equation}
where $\gammalin$ is the linear instability growth rate.
Using \equ{eq:p6res} to compute $\psixperpirtilde$ and substituting
the result into \equ{eq:momentumiorder5perp}, the classical linear
mirror dispersion relation \citep{hellinger07} is recovered:
\begin{equation}
\label{eq:growthrate}
\gammalin\!=\!\sqrt{\f{2}{\pi}}|\kpar|\vtilde
\left[\Gamma_2-\f{3}{2}\,\rhotilde^2\kperp^2-\left(\f{\pzeroperpi-\pzeropari}{\pzeroperpi}+\f{2}{\betazeroperpi}\right)\f{\kpar^2}{\kperp^2}\right]~,
\end{equation}
with the effective thermal speed
\begin{equation}\vtilde^{-1}=-\sqrt{2\pi}\,\f{2
  m_i}{\pzeroperpi}\int\f{\vperp^4}{4}\left.\dpart{\fzeroi}{\vpar^2}\right|_{\vperp}\!\!\!\!\!\delta(\vpar)\,\diff^3\vec{v}~.
\end{equation}
Because of the resonance, $\ffourirtilde$ develops a velocity-space
boundary layer on the timescale $\sim 1/\gammalin$, resulting in
a correction to the line-averaged distribution function that satisfies:
\begin{equation}
\label{eq:driftkinlineav}
\dlag{\lineav{\ffourir}}{t}=-\mu \bo\,
\lineav{\left(\f{\gradpar{\bfourpartilde}}{\bo}\right)\dpart{\ffourir}{\vpar}}+C[\lineav{\ffourir}]~.
\end{equation}
Assuming a monochromatic perturbation for simplicity and using
\equ{eq:f4itilde} to calculate the line-averaged nonlinear term on the
r.h.s.~of \equ{eq:driftkinlineav}, we recover the resonant
quasilinear diffusion equation \citep{shapiro64}:
\begin{equation}
  \label{eq:quasilinear}
  \dpart{\lineav{\ffourir}}{t}\!=\!\dpart{}{\vpar}\left[\f{2\left(\mu\bo\right)^2\kpar^2\gammalin}{\gammalin^2+\left(\kpar\vpar\right)^2}\lineav{\left(\f{\bfourpartilde}{\bo}\right)^2}\dpart{\fzeroi}{\vpar}\right]+C[\lineav{\ffourir}]~.
\end{equation} 
The effect of the first term on the r.h.s.~of \equ{eq:quasilinear} is 
to relax $\Gamma_2$ (see \equ{eq:Gamma2}) by flattening the total
averaged distribution function at low $\xi$, thereby decreasing the
growth rate \cite{pokhotelov08,hellinger09}.

\paragraph{Trapping regime.}
Quasilinear relaxation ceases to be the dominant saturation mechanism
once particle trapping becomes dynamically significant
($\omegab\sim \kpar\vthi\,(\dB/B)^{1/2}\sim \gammalin$). 
Indeed, due to the growth of $\dB/B$ and quasilinear
reduction of the growth rate $\dpartshort{}{t}\ll \gammalin$ for
$t\gg 1/\gammalin$, (i) the system eventually reaches a
bounce-dominated regime, $\omegab \gg \dpartshort{}{t}$, and (ii) 
collisional and shearing effects, however slow their timescales are,
compared to the initial linear instability timescale,
inevitably become important after a few instability times
(hence the maximal ordering
\equs{eq:expansionsummaryfirst}{eq:expansionsummarylast}). To elicit
these effects, we rewrite \equ{eq:driftkin} in $(\mu,E=v^2/2)$ variables:
\begin{equation}
  \label{eq:driftkinEmu}
  \dlag{f_i}{t}\pm\vparEmu\,\dpart{f_i}{\ell}= -\mu\dlag{B}{t}\dpart{f_i}{E}+C[f_i]~,
\end{equation}
where $\ell$ is the distance along the perturbed field line. Expanding
\equ{eq:induction} and \equ{eq:driftkinEmu} at $\bigO{\eps^6}$, we
obtain
\begin{eqnarray}
  \label{eq:driftkinexpandEmu}
  \dlag{\ffourir}{t}\pm\vparEmu\,\dpart{\ffourir}{\ell} & = &
  -\mu\bo\dlag{}{t}\f{\bfourpar}{\bo}\dpart{\fzeroi}{E} \\
& &
\hspace{-2.5cm} -\mu\bo\left(\dlag{\ln\bo}{t} +\dlag{}{t}\f{\bfourpar}{\bo}\right)\dpart{\ffourir}{E} + C[\ffouri]~.\nonumber
\end{eqnarray}
Here $C[\fzeroi]$ and
$-\mu\bo\,\left(\dlagshort{\ln\bo}{t}\right)
(\dpartshort{\fzeroi}{E})$ have been
 discarded for the same reason as in \equ{eq:driftkinexpand}. Note
that both $\dpartshort{\ffourir}{E}$ and $\dpartshort{\ffourir}{\mu}$
are $\bigO{1}$ because of the velocity-space boundary layer in
$|\xi|<\xitr=\bigO{\eps^2}$.

We anticipate that magnetic fluctuations will grow secularly as
$\bfourpar\sim\bfourpar (t_{\lin})(t/t_{\lin})^s$, with $s>0$, 
for $t\gg t_{\lin} \sim 1/\gammalin\,(\sim 1/\omegab)$, and introduce
a secondary ordering parameter $\chi=(t_\lin/t)^{s/2}\ll 1$,
so now $\bfourpar/\bo=\bigO{\eps^4/\chi^2}$ and $\xitr\sim
\left(\bfourpar/\bo\right)^{1/2}= \bigO{\eps^2/\chi}\gg\eps^2$.
The instantaneous nonlinear growth rate is $\gammanl\sim
\dpartshort{}{t}\sim 1/t=\bigO{\eps^2\chi^{2/s}}$,
so the new ordering guarantees
$\omegab\sim\kpar \vthi\,\xitr\gg \gammanl$.
For trapped particles to play a role in the nonlinear evolution, their
pressure in \equ{eq:momentumiorder5perp}
must be taken to be of the same order as the instability-driving term,
$\Gamma_2\,(\bfourpar/\bo) \sim \psixperpir/\pzeroperpi = \bigO{\eps^2/\chi^2}$.
Given that $\psixperpir\sim\xitr\ffourir$, we must therefore order
$\ffourir=\bigO{\eps^4/\chi}$. Expanding \equ{eq:driftkinexpandEmu}
to lowest order $\bigO{\eps^6/\chi^2}$, we find that the
distribution function of the trapped particles is
homogenized along the field lines within the traps:
$\dpartshort{\ffourir}{\ell}=0$. Therefore,
$\ffourir=\bounceav{\ffourir}+{\ffourir}'$, where
${\ffourir}'\ll\bounceav{\ffourir}$ and
$\bounceav{\bullet}=\oint\,\bullet\, \diff\ell$ denotes a bounce
average between bounce points $\ell_1$ and $\ell_2$ defined by the relation
$E=\mu B(\ell_1)=\mu B(\ell_2)$. Looking at the next orders
of \equ{eq:driftkinexpandEmu}, we find that the first term on the
r.h.s.~(the betatron term linear in perturbations) is
$\bigO{\eps^6\chi^{2/s-2}}$, while the time derivative on the
l.h.s.~and the r.h.s.~term quadratic in perturbations are
$\bigO{\eps^6\chi^{2/s-1}}$, so quasilinear effects are
subdominant. The terms involving $\dlagshort{\ln \bo}{t}$ and
collisions are $\bigO{\eps^6\chi}$. For \equ{eq:driftkinexpandEmu} to
have a solution at $\bigO{\eps^6\chi}$,  we see that $s=2/3$ is
required, so~$\bfourpar/\bo \propto t^{2/3}$. The resulting equation
for ${\ffourir}'$ is
\begin{eqnarray}
\pm\dpart{{\ffourir}'}{\ell} &\! = & 
\!- \f{\mu\bo}{\vparEmu}\left(\dlag{}{t}\f{\bfourpar}{\bo}\dpart{\fzeroi}{E}\!+\!
\dlag{\ln\bo}{t} \dpart{\,\bounceav{\ffourir}}{E}\right)\nonumber\\ &
&\!+C[\bounceav{\ffourir}]~.
\label{eq:deltaequation}
\end{eqnarray}
Using a Lorentz operator and bounce averaging, we obtain
\begin{eqnarray}
\bounceav{\f{\mu\bo}{\vparEmu}\dlag{}{t}\f{\bfourpar}{\bo}}\dpart{\fzeroi}{E}
  & = & \label{eq:trappedequation}
\\
&& \hspace{-2cm}
- \dlag{\ln\bo}{t} \bounceav{\f{\mu\bo}{\vparEmu}}\dpart{\,\bounceav{\ffourir}}{E}\nonumber
\\
 & &\hspace{-2cm} +\f{\nuii}{\bo}\dpart{}{\mu}\left(\mu\,\bounceav{\vparEmu}\dpart{\,\bounceav{\ffourir}}{\mu}\right)~.\nonumber
\end{eqnarray}
\paragraph{Physical behavior and temporal evolution.}
This equation has taken some effort to derive but is fairly
transparent physically. It represents a competition between
perpendicular betatron cooling of the equilibrium distribution due to
the local
decrease of the magnetic field in the deepening mirror traps
(the l.h.s.~of \equ{eq:trappedequation}), the perpendicular 
betatron heating of the trapped-particle population associated
with the increasing mean field $\bo$ (the first term on the r.h.s.),
and their collisional isotropization (the second term on
the r.h.s.). In the weakly collisional, unsheared regime ($\nuii\neq 0$,
$\dlagshort{\ln\bo}{t} \equiv S = 0$), the balance is between betatron
cooling and collisions. In the collisionless, shearing
regime ($\nuii=0$, $S > 0$), it is instead between
betatron cooling (of the bulk distribution in mirror perturbations) and
heating (of the perturbed distribution in the growing mean field). 
In order for the system to stay marginal in the face of continued driving
and/or collisional relaxation, magnetic perturbations have to continue
growing. A simple physical interpretation of the collisionless case is that
trapped particles regulate the evolution so as to ``see'' effectively
a constant total magnetic field. 

Solutions of \equsthree{eq:momentumiorder5perp}{eq:p6res}{eq:trappedequation}
can be found in the form $\bfourpar/\bo=\mathcal{A}(t)\mathcal{B}(\vec{\ell})$. Using
\equ{eq:trappedequation}, this implies
$\bounceav{\ffourir}=\alpha\,\mathcal{A}\,\left(\dlagshort{\mathcal{A}}{t}\right)\,\mathcal{H}[(E-\mu
\bo)/\mathcal{A}(t)]\,\dpartshort{\fzeroi}{E}$, where $\alpha=1/S$ if
$\nuii=0$ and $1/\nuii$ if $S=0$ ($\mathcal{H}$ also depends on
the functional form of $\mathcal{B}(\vec{\ell})$). Then, from \equ{eq:p6res},
$\psixperpir=\alpha\,
\mathcal{A}^{3/2}\left(\dlagshort{\mathcal{A}}{t}\right)\,\mathcal{F}(\vec{\ell})$.
\Equ{eq:momentumiorder5perp} will have solutions if $\alpha\, \mathcal{A}^{1/2}
\left(\dlagshort{\mathcal{A}}{t}\right) = \Lambda \Gamma_2$, with $\Lambda$
a constant of order unity. As anticipated in our discussion of the
secondary ordering (in $\chi$), this implies that perturbations grow
secularly as
\begin{equation}
\mathcal{A}(t) = (\Lambda \Gamma_2 S t)^{2/3} \ \mathrm{and}\
\mathcal{A}(t) = (\Lambda\Gamma_2\nuii t)^{2/3}
\end{equation}
in the shearing-collisionless regime and unsheared-collisional regime,
respectively. This result is formally valid for times $St, \nuii\, t
=\bigO{\eps^4}$ \footnote{These results do not apply to the frequently discussed
``idealized'' case $\nuii=S=0$, because continued growth of 
fluctuations was assumed to derive \equ{eq:trappedequation}. 
In that regime, fluctuations should instead settle in a
steady state $\dB/B \sim \Gamma^2$ through quasilinear 
relaxation, possibly after a few transient bounce oscillations
\citep{istomin09}.
However, only a small amount of collisions or continued
shearing/compression is required for the
present theory to apply: these effects are bound to become dynamically
important after a few instability times, once quasilinear relaxation
has reduced the instability drive $\Gamma$ sufficiently.}. 
The $t^{2/3}$ time dependence also holds in mixed
regimes ($\nuii\neq 0$, $S\neq 0$). $\Lambda$ and
$\mathcal{B}(\vec{\ell})$ (which is not sinusoidal) are obtained by
solving a nonlinear eigenvalue equation involving trapping
integrals. A detailed classification of the solutions of this equation
lies beyond the scope of this Letter.

\paragraph{Conclusion.} 
\Equ{eq:trappedequation} and the secular growth of the nonlinear mirror
instability, $\dB/B\propto t^{2/3}$, in both collisional and
collisionless regimes, are the main results of this work. 
Thus, we appear to be approaching a theory in which trapping effects,
higher amplitude nonlinearities
\citep{kuznetsov07,califano08,istomin09,pokhotelov10} 
and relaxation of anisotropy through anomalous particle scattering
\citep{kunz14a} blend together harmoniously.
The results make manifest the importance of particle trapping
\cite{kivelson96,pantellini98}. Numerical simulations \citep{kunz14a}
confirm $t^{2/3}$ secular growth of mirror perturbations in a
collisionless, shearing plasma, with saturation amplitudes
$\dB/B=\bigO{1}$ independent of $S$ (cf. \citep{riquelme14}). 

The weakly collisional, weakly shearing regimes studied in this
Letter occur in many natural environments
\citep{quataert01,bale09,rosin11} and are increasingly the
focus of attention in the context of high-energy astrophysical plasmas
\citep{schekochihin08,rosin11,kunz14a,riquelme14}. The emergence
of finite-amplitude magnetic mirrors with scales smaller than the mean
free path lends credence to the idea that microscale instabilities
regulate processes such as heat conduction \citep{chandran98},
viscosity, heating \citep{sharma06,sharma07,kunz11} and dynamo
\citep{schekochihin06,mogavero14} in such plasmas, and therefore 
profoundly alter their large-scale energetics and dynamics.

\paragraph{Acknowledgements.} The authors thank I. Abel, G.~Hammett,
R.~Kulsrud, M.~Kunz and T.~Passot for many helpful discussions and suggestions.

\newcommand{\etal}{\textit{et al.}}
\bibliographystyle{apsrev}

\end{document}